\documentclass[twocolumn,showpacs,preprintnumbers,superscriptaddress,amsmath,amssymb]{revtex4}
\usepackage{graphicx}% Include figure files
\usepackage{dcolumn}% Align table columns on decimal point
\usepackage{bm}% bold math
\usepackage{amssymb}
\usepackage{amssymb}
\usepackage{verbatim}
\usepackage{color}
\def\>{\rangle}

\begin{document}

\title{Entanglement distribution over quantum code-division-multiple-access networks}% Force line breaks with\\
\author{Chang-long Zhu}
\affiliation{Department of Automation, Tsinghua University,
Beijing 100084, P. R. China} \affiliation{Center for Quantum
Information Science and Technology, TNList, Beijing 100084, P. R.
China}
\author{Nan Yang}
\affiliation{Department of Automation, Tsinghua University,
Beijing 100084, P. R. China} \affiliation{Center for Quantum
Information Science and Technology, TNList, Beijing 100084, P. R.
China}
\author{Yu-xi Liu}
\affiliation{Institute of Microelectronics, Tsinghua University,
Beijing 100084, P. R. China} \affiliation{Center for Quantum
Information Science and Technology, TNList, Beijing 100084, P. R.
China}
\author{Franco Nori}
\affiliation{CEMS, RIKEN, Saitama 351-0198, Japan}
\affiliation{Physics Department, The University of Michigan, Ann
Arbor, Michigan 48109-1040, USA}
\author{Jing Zhang}\email{jing-zhang@mail.tsinghua.edu.cn}
\affiliation{Department of Automation, Tsinghua University,
Beijing 100084, P. R. China} \affiliation{Center for Quantum
Information Science and Technology, TNList, Beijing 100084, P. R.
China}

\date{\today}

\begin{abstract}
We present a method for quantum entanglement distribution over a
so-called code-division-multiple-access network, in which two
pairs of users share the same quantum channel to transmit
information. The main idea of this method is to use different
broad-band chaotic phase shifts, generated by electro-optic
modulators (EOMs) and chaotic Colpitts circuits, to encode the
information-bearing quantum signals coming from different users,
and then recover the masked quantum signals at the receiver side
by imposing opposite chaotic phase shifts. The chaotic phase
shifts given to different pairs of users are almost uncorrelated
due to the randomness of chaos and thus the quantum signals from
different pair of users can be distinguished even when they are
sent via the same quantum channel. It is shown that two maximally-entangled
states can be generated between two pairs of users by
our method mediated by bright coherent lights, which can be more
easily implemented in experiments compared with single-photon
lights. Our method is robust under the channel noises if only the
decay rates of the information-bearing fields induced by the
channel noises are not quite high. Our study opens up new
perspectives for addressing and transmitting quantum information
in future quantum networks.

\end{abstract}


\maketitle

\section{Introduction}\label{s1}
With recent progresses in various quantum systems, such as
ion-trap systems~\cite{Lukin,LMDuan,Sangouard} and solid-state
quantum circuits~\cite{You,YMakhlin,Xiang}, it is now possible to
discuss how to establish more efficient quantum networks or
so-called quantum internet~\cite{Kimble}. The existing studies
about quantum communication~\cite{Sillanpaa,Wang,Pan} and quantum
cryptography~\cite{Gisin,Razavi} have shown that quantum network
has great advantages to transfer classical or quantum information.
However, how to best transfer information via quantum networks is
still an open
problem~\cite{Mabuchi1,Maitre,Phillips,LMDuan2,Matsukevich,Acin,XinYou2,Felinto,vanderWal,Chou,Wilk}.

To transfer quantum information over a large-scale quantum
network, a question is: can we allow different pairs of users, who
want to transmit information, to share the same
channel~\cite{Yard,Horodecki1,Horodecki2}. This problem has been
widely discussed in the field of classical
communication~\cite{Rom,Ipatov}. In classical communication
systems, such kind of methods are called channel-access methods or
multiple-access methods. There are mainly four different kinds of
multiple-access methods~\cite{Frenzel}: the frequency-division
multiple access (FDMA) methods, the time-division multiple access
(TDMA) methods, the code-division multiple access (CDMA) methods,
and the orthogonal-frequency division multiple access (OFDMA)
methods. In FDMA methods, different frequency bands are assigned
to different data streams, while in TDMA methods the users split
their signals into pieces and transmitted them at different time
slots to share the same channel. TDMA and FDMA work equally well
and are the key techniques for the first generation (analog) and
the second generation (digital) mobile networks. In CDMA methods,
each pair of users shares the same channel and distinguishes with
each other by their own unique codes. It can be shown that CDMA
can accommodate more bits per channel use, compared with TDMA and
FDMA~\cite{TMCover}, and thus is used in third-generation mobile
communication systems. However, the interference between different
data streams will deteriorate the information rate of CDMA. Other
competitive approaches are proposed including orthogonal
frequency-division multiple-access (OFDMA), in which the available
subcarriers are divided into several mutually orthogonal
subchannels which are assigned to distinct users for simultaneous
transmission. The OFDMA is capable of avoiding the interference
problem and thus provide better performance in classical {\it
digital} communication.

Although the multiple-access problem has been widely studied in
classical communication, it is considered in quantum communication
only recently due to the development of techniques for scalable
quantum network. To our knowledge, FDMA, or equivalently, the
so-called wavelength-division multiple access (WDMA), has been
used for quantum key distribution
(QKD)~\cite{Godbout,Townsend,Yoshino,Brassard1,Brassard2,IChoiOE:2010,JRoslundNatPhonon:2014},
in which classical information is transmitted over quantum
network, and TDMA has been used to generate large entangled
cluster states~\cite{AFurusawaNatPhotonics:2013}. However, whether
more popular classical communication technique such as
CDMA~\cite{JZhang3,Carcia,Humble,Carcia2} and
OFDMA~\cite{Fei,Anandan} can be applied to quantum communication
systems is still an interesting problem yet to be solved.

Recently, various protocols are proposed to extend CDMA to the
quantum case~\cite{JZhang3,Carcia,Humble,Carcia2} and there is
evidence showing that CDMA can provide higher information rates
for quantum communication compared with FDMA~\cite{JZhang3}. In
Ref.~\cite{JZhang3}, particular chaotic phase shifts, which work
as the unique code in CDMA, are introduced to spread the
information-bearing quantum signals in the frequency regime. Since
the chaotic phase shifts introduced for different users are
uncorrelated, the cryptographic quantum signals from different
users are orthogonal and thus can be distinguished even when we
transmit them via the same channel. The cryptographic quantum
signals can be decoded by introducing reversed chaotic phase
shifts at the receiver side, by which the transmitted quantum
information can be recovered coherently. The physical media used
in Ref.~\cite{JZhang3} to transmit quantum information are
single-photon lights~\cite{Humble}.

Different from the protocol in Ref.~\cite{JZhang3}, instead of
single-photon lights, we use bright coherent lights~\cite{Carcia2}
to transmit quantum information over quantum CDMA network, which
are easier to be realized in experiments. We find that quantum
entanglement can be controllably distributed between two pairs of
users sharing a single quantum channel. We also present the
particular design of the chaotic phase shifters used in our
proposals by introducing electro-optic modulators (EOMs) and
chaotic Colpitts oscillator circuits~\cite{Kennedy} which is not
clearly discussed in Ref.~\cite{JZhang3}. The Pecora-Carrol
synchronization technique~\cite{Shizhiguo,Parlitz} is introduced
to generate the reverse chaotic phase shifts at the receiver side.
This paper is organized as follows. In Sec.~\ref{s2}, we present
the general description of the quantum CDMA network, we use to
transmit quantum information. In Sec.~\ref{s3}, we state how to
distribute maximally-entangled quantum states over proposed
quantum CDMA network mediated by bright coherent lights. In
Sec.~\ref{s4}, we consider the non-ideal case to see how channel
noise will affect our main results. In Sec.~\ref{s5}, we present
the conclusions and a forecast of future work.

\section{QUANTUM CDMA NETWORK BY CHAOTIC SYNCHRONIZATION}\label{s2}

The main purpose of our work is to generate two maximally-entangled
states between two pairs of nodes (one pair of nodes are
node $1$ and node $3$ and the second pair of nodes are node $2$
and node $4$) via a single quantum channel (see Fig.~\ref{Fig of
the schematic diagram of quantum multiple access network}). The
quantum signals sent by node $1$ and node $2$ are first encoded by
two chaotic phase shifters ${\rm CPS}_1$ and ${\rm CPS}_2$, and
the two output beams are combined by a $50\!:\!50$ beam splitter and
then transmitted via a quantum channel. At the receiver side, this
combined quantum signal is divided into two branches by another
$50\!:\!50$ beam splitter and sent to another two chaotic phase
shifters ${\rm CPS}_3$ and ${\rm CPS}_4$ introduced to decode the
information. The recovered quantum signals are then sent to the
two receiver nodes.

\begin{figure}[h]
\centerline{
\includegraphics[width=8.4 cm, clip]{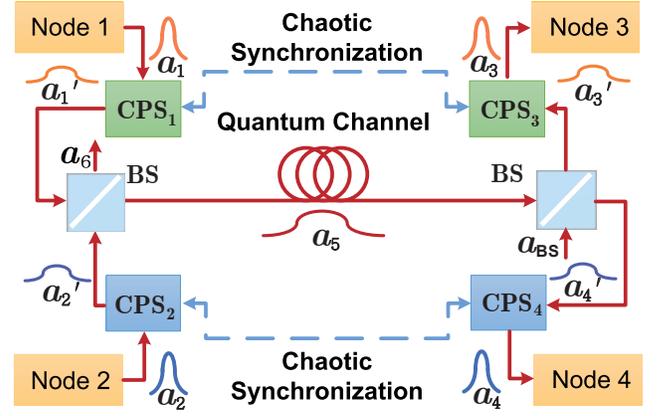}}
\caption{(color online) Schematic diagram of the quantum CDMA
network by chaotic synchronization. The wave packets are broadened
by two chaotic phase shifters, i.e., ${\rm CPS}_1$ and ${\rm
CPS}_2$, at the senders, and then recovered by another two chaotic
phase shifters, i.e., ${\rm CPS}_3$ and ${\rm CPS}_4$, at the
receiver side.}\label{Fig of the schematic diagram of quantum
multiple access network}
\end{figure}

To understand the encoding and decoding processes of our method,
let us assume that the optical field entering the $i$-th chaotic
phase shifter is $a_i$ ($i=1,2,3,4$). The chaotic phase shifter
${\rm CPS}_1$ (${\rm CPS}_2$) induces an effective Hamiltonian
$\delta_1\left(t\right)a^{\dagger}_1 a_1$
$[\delta_2\left(t\right)a^{\dagger}_2 a_2]$, where
$\delta_1\left(t\right)$ $[\delta_2\left(t\right)]$ is a classical
chaotic signal. It can be shown that ${\rm CPS}_1$ (${\rm CPS}_2$)
leads to phase-shift factor
$\exp\left(-i\theta_1\left(t\right)\right)$
$[\exp\left(-i\theta_2\left(t\right)\right)]$ for the optical
field. At the receiver side, the chaotic phase shifter ${\rm
CPS}_3$ (${\rm CPS}_4$) induces an opposite Hamiltonian
$-\delta_1\left(t\right)a^{\dagger}_3 a_3$
$[-\delta_2\left(t\right)a^{\dagger}_4 a_4]$ by which a reversed
phase-shift factor $\exp\left(i\theta_1\left(t\right)\right)$
$[\exp\left( i \theta_2 \left( t \right) \right]$ is introduced
to decode the information-bear signal masked by the chaotic phase
shift. Here $\theta_i=\int_0^t\delta_i\left(t\right)dt$, $i=1,2$.
To assure that the chaotic phase shift at the sender side and that
at the receiver side can be exactly cancelled, an auxiliary
classical channel between node $1$ (node $2$) and node $3$ (node
$4$) is introduced to synchronize the two chaotic phase shifters
\cite{Pecora,Cuomo} (see Fig.~\ref{Fig of the schematic diagram of
quantum multiple access network}).

The whole information transmission process can be represented by
the input-output relationship of the whole quantum network from
$a_1$, $a_2$ to $a_3$, $a_4$ (see Fig.~\ref{Fig of the schematic
diagram of quantum multiple access network}). To derive it, we can
see that the input-output response of the chaotic phase shifters
${\rm CPS}_i, \,i=1,2,3,4$ can be written as
\begin{eqnarray}\label{Input-output relationship of the chaotic phase shifters}
&a_1'=a_1 e^{-i\theta_1}, \quad a_2'=a_2 e^{-i\theta_2},&
\nonumber\\
&a_3=a_3' e^{i\theta_1}, \quad a_4=a_4' e^{i\theta_2},&
\end{eqnarray}
and the input-output response of the two beam splitters ${\rm
BS}_1$ and ${\rm BS}_2$ can be written as follows:
\\[0.05cm]

\noindent ${\rm
BS}_1$:
\begin{equation}\label{Input-output relationship of BS1}
a_5=\frac{1}{\sqrt{2}}a_1'+\frac{1}{\sqrt{2}}a_2',\quad
a_6=\frac{1}{\sqrt{2}}a_1'-\frac{1}{\sqrt{2}}a_2',
\end{equation}

\noindent ${\rm BS}_2$:
\begin{equation}\label{Input-output relationship of BS2}
a_3'=\frac{1}{\sqrt{2}}a_5+\frac{1}{\sqrt{2}}a_{\rm BS},\quad
a_4'=\frac{1}{\sqrt{2}}a_5-\frac{1}{\sqrt{2}}a_{\rm BS},
\end{equation}

Hence, from Eqs.~(\ref{Input-output relationship of the chaotic phase
shifters}-\ref{Input-output relationship of BS2}),
we can obtain the input-output relationship of the whole quantum network

\begin{eqnarray}\label{Input-output relationship between the senders and receivers}
a_3&=&\frac{1}{2}a_1+\frac{1}{2}a_2 e^{i\left(\theta_1-\theta_2\right)}+
\frac{1}{\sqrt{2}}e^{i\theta_1}a_{\rm BS},\nonumber\\
a_4&=&\frac{1}{2}a_2+\frac{1}{2}a_1
e^{i\left(\theta_2-\theta_1\right)}-
\frac{1}{\sqrt{2}}e^{i\theta_2}a_{\rm BS}.
\end{eqnarray}

For the chaotic phase shifts $\theta_i\left(t\right),\,i=1,2$, we
should take average over these broadband ``random"
phases~\cite{JZhang}, by which we have: $\overline {\exp ( \pm
i{\theta _i}(t))}  \approx \sqrt
{{M_i}}$~\cite{Zhou,Ashhab,Bergli}, where
\begin{equation}\label{Chaotic correction factor}
M_i=\exp\left[-\pi\int_{\omega_{li}}^{\omega_{ui}} d\omega
S_{\delta_i}\left(\omega\right)/\omega^2\right].
\end{equation}
$S_{\delta_i}\left(\omega\right)$ is the power spectrum density of
the signal $\delta_i\left(t\right)$ and $\omega_{li}$,
$\omega_{ui}$ are the ``lower" and ``upper" bounds of the
frequency band of $\delta_i\left(t\right)$, respectively.
Equation~(\ref{Input-output relationship between the senders and
receivers}) can then be reduced to
\begin{eqnarray}\label{Input-output relationship between the senders and receiver: reduction}
a_3&=&\frac{1}{2}a_1+\frac{\sqrt{M_1M_2}}{2}a_2+\sqrt{\frac{M_1}{2}}a_{\rm BS},\nonumber\\
a_4&=&\frac{1}{2}a_2+\frac{\sqrt{M_1M_2}}{2}a_1-\sqrt{\frac{M_2}{2}}a_{\rm BS}.
\end{eqnarray}
The correction factor $M_i$ may become extremely small when
$\delta_i\left(t\right)$ is induced by a chaotic signal which has
a broadband frequency spectrum. Thus we have $a_3 \approx a_1/2$,
$a_4 \approx a_2/2$, which means that the quantum signal
transmitted from node $1$ to node $3$ and the quantum signal
transmitted from node $2$ to node $4$ can be totally decoupled
from each other although they are transmitted simultaneously on
the same quantum channel. The mechanism of such a quantum multiple
access network is that the information bearing fields transmitted
on the quantum channel are broadened by the chaotic phase shifters
in the frequency regime, which cannot be detected unless we can
reduce the chaotic phase shifts and sharpen the quantum signal by
chaotic synchronization. This idea is quite similar to the
classical CDMA communication. That is why we call it quantum CDMA
network in Ref.~\cite{JZhang3}.

Now, let us consider the case where the quantum fields $a_1$ and
$a_2$ are in the coherent states $| \alpha_1 \rangle$ and $|
\alpha_2 \rangle$, and the field $a_{\rm BS}$ is in a vacuum
state. It can be easily checked from the input-output relationship
given by Eq.~(\ref{Input-output relationship between the senders
and receiver: reduction}) that the output fields $a_3$ and $a_4$
of the quantum network are in the coherent states
\begin{eqnarray}\label{Output state of the quantum CDMA network under coherent light}
|\alpha_3\rangle &=& \left|\frac{1}{2}\alpha_1+\frac{1}{2}\sqrt{M_1M_2} \alpha_2 \right\rangle\approx\left|\frac{1}{2}\alpha_1\right\rangle, \nonumber\\
|\alpha_4\rangle &=&
\left|\frac{1}{2}\alpha_2+\frac{1}{2}\sqrt{M_1M_2} \alpha_1
\right\rangle\approx\left|\frac{1}{2}\alpha_2\right\rangle.
\end{eqnarray}

\section{QUANTUM ENTANGLEMENT DISTRIBUTION OVER Q-CDMA NETWORK}\label{s3}

Let us then consider how to distribute two-qubit quantum
entanglement over the quantum CDMA network. In our proposal, the
qubit states are stored in the dark states of four $\Lambda$-type
three-level atoms in four optical cavities (see
Fig.~\ref{Schematic diagram of quantum state transmission over
quantum multiple access network by chaotic synchronization}). What
we want to do is to generate maximally-entangled state between
atom $1$ (atom $2$) and atom $3$ (atom $4$). Here we extend the
strategy in Refs.~\cite{PvanLoock,Ladd} to generate such kinds of
distributed entangled states by bright coherent lights. The
Hamiltonian of the $i$-th coupled atom-cavity system can be
expressed as
\begin{equation}\label{JC model}
\tilde{H}_i^{qc}=\omega_c a_i^{\dagger} a_i +\frac{\omega_q}{2}
\sigma_{z}^{(i)} + g \left( a_i^{\dagger} \sigma_{-}^{(i)} + a_i \sigma_{+}^{(i)}\right),
\end{equation}
where $\omega_c$, $a_i$ ($a_i^{\dagger}$) are the frequency and
the annihilation (creation) operator of the cavity mode;
$\omega_q$, $\sigma_{z}^{(i)}$ and $\sigma_{\pm}^{(i)}$ are the
frequency, the z-axis Pauli operator, and the ladder operators of
the qubit; and $g$ is the coupling strength between the qubit and
the cavity mode. Here, to simplify the discussion, we have assumed
that the system parameters are the same for four qubit-cavity
systems. Under the dispersive-detuning condition
$|\Delta|=|\omega_c-\omega_q|\gg|g|$, the Hamiltonian can be
diagonalized and reexpressed in the interaction picture
as~\cite{Blais2}
\begin{equation}\label{Hamitonian of the coupled atom-cavity system in the dispersive regime}
H_i^{qc}=\frac{g^2}{\Delta} a_i^{\dagger} a_i \sigma_{z}^{(i)}.
\end{equation}

\begin{figure*}
\centerline{\includegraphics[width=18 cm, clip]{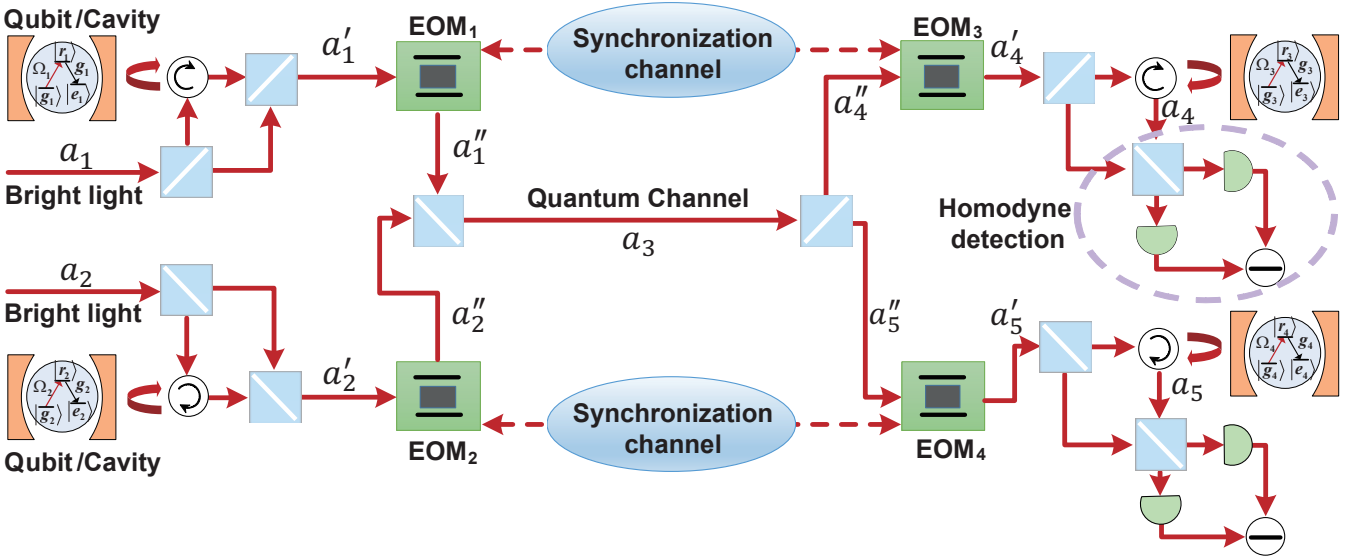}}
\caption{(color online) Schematic diagram of the entanglement
distribution over a quantum multiple-access network. The bright
coherent lights are sent to two cavities $1$ and $2$ in which the
optical fields interact with the atoms $1$ and $2$, respectively.
After that, the two output beams transmit through two ${\rm
EOM}$s, i.e., $\rm{EOM}_1$ and $\rm{EOM}_2$, and are broadened in
the frequency domain. The two output beams are then combined by an
beam splitter and transmitted through a single quantum channel. At
the receiver side, the combined optical field is split into two
branches by another beam splitter and fed into another two ${\rm
EOM}$s, i.e., ${\rm EOM}_3$ and ${\rm EOM}_4$. Since the chaotic
circuits driving $\rm{EOM}_1$~($\rm{EOM}_2$) and
$\rm{EOM}_3$~($\rm{EOM}_4$) are synchronized, the chaotic phases
introduced at the sender side can be compensated at the receiver
side, and thus the quantum signals transmitted can be recovered.
The recovered quantum signals are then stored in the dark states
of the atoms $3$ and $4$, and homodyne detections are preformed
for the output fields to post-select the maximally entangled
states. }\label{Schematic diagram of quantum state transmission
over quantum multiple access network by chaotic synchronization}
\end{figure*}

In this paper, we introduce four electro-optic modulators
(EOMs)~\cite{Kolchin} acting as the chaotic phase shifters ${\rm
CPS}_i$. It is known that the refractive index of the
electro-optic crystal in EOM can be varied by changing the voltage
${V\!\left(t\right)}$ acting on it (see Fig.~\ref{A transverse
electro-optical modulator and schematic diagram of the chaotic
signal generantion and synchronization device}(a)). Based on this
effect, we let the information-bearing optical field pass through
the EOM to obtain a phase shift $\beta$ which will be changed by
varying the voltage ${ V\!\left(t\right)}$ acting on it. This
phase shift can be expressed as $\,\beta = \left[\left(\omega n^3
r L\right)/\left(c d\right)\right]\,V\!\left(t\right)$, where
$\omega$ is the frequency of the injected light. $n$ are the
refractive index and the electro-optic coefficient of the
electro-optic crystal in EOM. $L$ and $d$ are respectively the
length and thickness of the EOM (see Fig.~\ref{A transverse
electro-optical modulator and schematic diagram of the chaotic
signal generantion and synchronization device}(a)). $c$ is the
velocity of light. Therefore, when the optical field transmits
through the EOM, an interaction Hamiltonian
$H_i=\delta_i\left(t\right) a_i^{\dagger}a_i=-\left(\hbar
/\tau\right)\beta a^{\dagger}_ia_i$~\cite{Tsang} can be obtained,
where $a_i~(a^{\dagger}_i)$ is the annihilation (creation)
operator of the injected field and $\tau$ is the optical
round-trip time through the EOM. In present system, each pair of
EOMs is driven by two synchronized standard chaotic Colpitts
oscillator circuits, as shown in Fig.~\ref{A transverse
electro-optical modulator and schematic diagram of the chaotic
signal generantion and synchronization device}(b), and the
specific synchronized circuit is presented in Appendix. We use the
voltage $V_{C2}$ to drive the EOM at the sender side, and the
voltage $\tilde{V}_{C2}$ to drive another EOM at the receiver
side, as shown in Fig.~\ref{A transverse electro-optical modulator
and schematic diagram of the chaotic signal generantion and
synchronization device}(b).

\begin{figure}[h]
\centerline{\includegraphics[width=8.6 cm, clip]{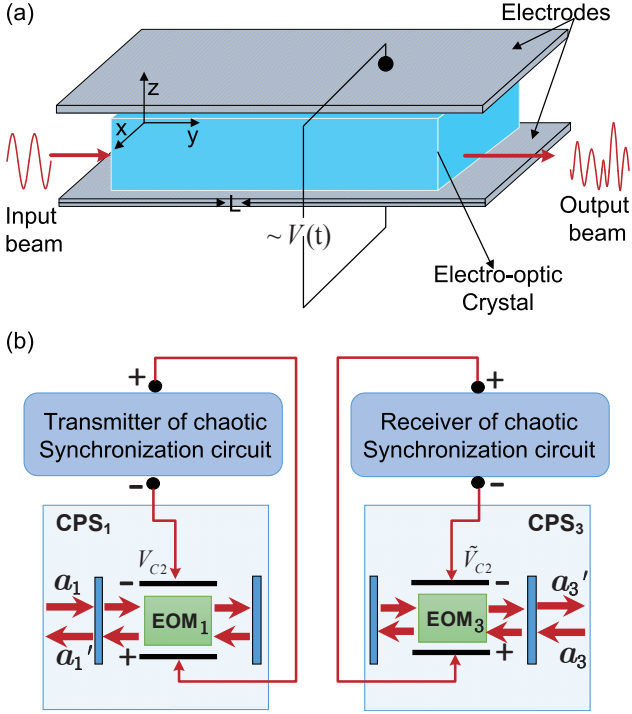}}
\caption{(color online) (a) Schematic diagram of a transverse
electro-optic modulator. The voltage is applied perpendicular to
the propagational direction of the input beam, and the refractive
index of the electro-optic crystal can be changed by varying the
voltage $V$, which induces a voltage-dependent phase shift on the
input beam; (b) The diagram of the chaotic synchronization circuit
between ${\rm CPS}_1$ and ${\rm CPS}_3$, where the transmitter
drives the ${\rm EOM}_1$ and the receiver drives the ${\rm
EOM}_3$.}\label{A transverse electro-optical modulator and
schematic diagram of the chaotic signal generantion and
synchronization device}
\end{figure}

To show how the quantum entanglement is distributed over our
quantum CDMA network, we assume that the $i$-th atom is in a
superposition state $|\psi_i \rangle=\left( |g_i \rangle + |e_i
\rangle \right)/\sqrt{2}$ ($i=1,2,3,4$). The probe field entering
the cavity $1$ (cavity $2$) is a bright coherent light
$|\alpha\rangle$ with average photon number $\bar{n}=|\alpha|^2\gg
1$. When the probe field comes out of cavity $1$ (cavity $2$)
at time $\tau'$, the system composed of the atom $1$ and the probe
field fed out of cavity $1$ is in an entangled state
\begin{equation}\label{entangled state of the atom 1 and the probe light 1}
e^{-i H_1^{qc} \tau'} |\psi_1\rangle |\alpha\rangle =
\frac{1}{\sqrt{2}} \left( |g_1\rangle |\alpha e^{-i\phi/2}\rangle
+ |e_1\rangle |\alpha e^{i\phi/2}\rangle \right).
\end{equation}
Similarly, the system composed of the atom $2$ and the probe light
fed out of cavity $2$ is also in an entangled state
\begin{equation}\label{entangled state of the atom 2 and the probe light 2}
e^{-i H_2^{qc} \tau'} |\psi_2\rangle |\alpha\rangle =
\frac{1}{\sqrt{2}} \left( |g_2\rangle |\alpha e^{-i\phi/2}\rangle
+ |e_2\rangle |\alpha e^{i\phi/2}\rangle \right).
\end{equation}
Here, $H^{qc}_1$, $H^{qc}_2$ are the Hamiltonians given by
Eq.~(\ref{Hamitonian of the coupled atom-cavity system in the
dispersive regime}) and $\phi = 2g^2\tau' /\Delta$ is the phase
shift of the probe field induced by the qubit-cavity coupling.
Thus, the system composed of atoms $1$, $2$ and the two probe
lights before entering our q-CDMA network is in a separable state
\begin{eqnarray*}
&\frac{1}{2}\left[ | g_1 g_2 \rangle | \alpha e^{-i \phi /2}
\rangle | \alpha e^{ -i \phi /2} \rangle+| e_1 g_2 \rangle |
\alpha e^{i \phi /2} \rangle | \alpha
e^{-i \phi /2} \rangle \right.\nonumber&\\
&\left.+|g_1 e_2\rangle | \alpha e^{-i \phi /2} \rangle | \alpha
e^{i \phi /2} \rangle+| e_1 e_2 \rangle | \alpha e^{i \phi /2}
\rangle | \alpha e^{i \phi /2} \rangle\right].&
\end{eqnarray*}
From Eq.~(\ref{Output state of the quantum CDMA network under
coherent light}), the system composed of the atoms $1$ and $2$ and
those two probe fields which enter the cavities $3$ and $4$ is in
the state
\begin{eqnarray*}
|\Phi \rangle & = & \left( \frac{1}{\sqrt{2}} | g_1 \rangle \left|
\frac{1}{2} \alpha e^{-i \phi /2} \right\rangle +
\frac{1}{\sqrt{2}} | e_1 \rangle \left| \frac{1}{2} \alpha e^{i
\phi /2} \right\rangle \right) \\
&& \left( \frac{1}{\sqrt{2}} | g_2 \rangle \left| \frac{1}{2}
\alpha e^{-i \phi /2} \right\rangle + \frac{1}{\sqrt{2}} | e_2
\rangle \left|\frac{1}{2} \alpha e^{i \phi /2} \right\rangle
\right).
\end{eqnarray*}
Here we have omitted those $\sqrt{M_1 M_2}$ terms since the
factors $M_1$ and $M_2$ are negligibly small in the chaotic
regime. After transmitting over the quantum CDMA network, the
probe fields $a_3$ and $a_4$ interact with the atoms $3$ and $4$,
and the interaction times are both $\tau'$. Thus, the state of the
total system composed of the four atoms and the optical fields fed
out of the quantum network is
\begin{eqnarray}\label{State of the total system}
\lefteqn{ e^{-i \left( H_3^{qc} + H_4^{qc} \right) \tau'} | \Phi
\rangle \frac{1}{2} \left( | g_3 \rangle +  | e_3 \rangle \right)
\left( | g_4 \rangle
+ | e_4 \rangle \right) } \nonumber \\
&& {\hspace{0.15cm}} = \left( \frac{1}{\sqrt{2}} | \Psi_{13}^{+}
\rangle \left| \frac{1}{2} \alpha \right\rangle + \frac{1}{2} |
g_1 g_3 \rangle
\left| \frac{1}{2} \alpha e^{-i \phi} \right\rangle \right. \nonumber \\
&& {\hspace{0.5cm}} \left. + \frac{1}{2} | e_1 e_3 \rangle \left|
\frac{1}{2} \alpha e^{i \phi}
\right\rangle \right) \nonumber \\
&& {\hspace{0.5cm}} \left( \frac{1}{\sqrt{2}} \left| \Psi_{24}^{+}
\right\rangle \left| \frac{1}{2} \alpha \right\rangle +\frac{1}{2}
| g_2 g_4 \rangle \left| \frac{1}{2}\alpha e^{-i \phi}
\right\rangle
\right. \nonumber \\
&& {\hspace{0.5cm}} \left. +\frac{1}{2} | e_2 e_4 \rangle \left|
\frac{1}{2} \alpha e^{i \phi} \right\rangle \right),
\end{eqnarray}
where $| \Psi^+_{13}\rangle = \left(|g_1 e_3\rangle + |e_1
g_3\rangle\right)/\sqrt{2}$ is the maximally-entangled state
between atom $1$ and atom $3$ and $| \Psi^+_{24}\rangle =
\left(|g_2 e_4\rangle + |e_2 g_4\rangle\right)/\sqrt{2}$ is the
maximally-entangled state between atom $2$ and atom $4$.

Finally, we impose homodyne detections on the probe fields leaking
out of the cavities $3$ and $4$. As shown in Eq.~(\ref{State of
the total system}), the state of the probe fields leaking out of
the cavities $3$ and $4$ can be three possible states $|
\alpha/2\rangle$, $| \alpha e^{-i\phi}/2\rangle$, and $| \alpha
e^{i\phi}/2\rangle$. Since the probe fields are bright coherent
lights with average photon number $\bar{n} = |\alpha|^2 \gg 1$, we
have
\begin{eqnarray*}
&& \left|\left\langle {\frac{1}{2} \alpha} \right. \left\vert {\frac{1}{2} \alpha
e^{-i \phi}} \right\rangle\right|^2 =
\exp\left[-\bar{n}\sin^2\left(\phi/2\right)\right]
\approx 0, \\
&& \left| \left\langle {\frac{1}{2} \alpha} \right. \left\vert {\frac{1}{2} \alpha
e^{i \phi}} \right\rangle \right|^2 = \exp\left[ -\bar{n}
\sin^2\left(
\phi/2 \right) \right] \approx 0, \\
&& \left| \left\langle {\frac{1}{2} \alpha e^{-i \phi}} \right. \left\vert
{\frac{1}{2} \alpha e^{i \phi}} \right\rangle \right|^2 = \exp\left(
-\bar{n}\sin^2\phi\right) \approx 0,
\end{eqnarray*}
which means that the three coherent states $|\alpha/2\rangle$,
$|\alpha e^{-i\phi}/2\rangle$, and $|\alpha e^{i\phi}/2\rangle$
are pairwise orthogonal and thus completely distinguishable. Thus,
the homodyne detections on the probe fields are just projective
measurements. Corresponding to the three measurement outputs
$\alpha/2$, $\alpha e^{-i\phi}/2$ and $\alpha e^{i\phi}/2$, the
states of the system composed of atom $1$ and $3$ (atom $2$ and
atom $4$) collapse to the maximally-entangled state $|
\Psi^+_{13}\rangle$ $\left(| \Psi^+_{24}\rangle\right)$ and two
separable states $| g_1 g_3\rangle$ $\left(| g_2
g_4\rangle\right)$ and $| e_1 e_3\rangle$ $\left(| e_2
e_4\rangle\right)$. The most important case is that the
measurement outputs of the probe fields leaking out of the cavity
$3$ and cavity $4$ are both $\alpha/2$. In this case, the atoms
$1$ and $3$ are in the maximally-entangled state $|
\Psi^+_{13}\rangle$ and the atoms $2$ and $4$ are in the maximally-entangled
state $| \Psi^+_{24}\rangle$, which means that we
generate two maximally-entangled states between two pairs of nodes
by sharing the same quantum channel.

We then consider the interference effects between the quantum
signals from the two pairs of users. These interference effects
have been omitted in our previous discussions under the condition
that the correction factor $M$~($M=M_1M_2$) is negligibly small if
the chaotic phases introduced have very broad bandwidths. However,
these interference effects will affect the information
transmission process if the bandwidths of the phase signals are
not broad enough. To show this, let us consider how the correction
factors $M_1$ and $M_2$ will change if we tune the correspondent
bandwidths of the chaotic signals $\delta_1$ and $\delta_2$. From
Fig.~\ref{Schematic of diagram of $M_1$, $M_2$ and $M_3$}(a), we
can see that both $M_1$ and $M_2$ decrease with the increase of
the bandwidth of $\delta_1$ and $\delta_2$, and when we take the
bandwidth values of the chaotic signals as $450$ MHz, $M_1$ and
$M_2$ are $0.0012$ and $0.0033$, respectively. With experimental
realizable parameters of the Colpitts chaotic
circuits~\cite{Shizhiguo,Parlitz}, it is not quite difficult to
generate a chaotic phase with bandwidth of $500$ MHz, and thus
$M_1$ and $M_2$ can be very small. In the meanwhile, we choose the
average photon number $\bar{n}=10$, and thus we have $M_1
M_2\ll4/\bar{n}$. This makes it reasonable to omit the
$\sqrt{M_1M_2}$ terms in Eqs.~(\ref{Input-output relationship
between the senders and receiver: reduction}), (\ref{Output state
of the quantum CDMA network under coherent light}), and
(\ref{Output state of the quantum CDMA network under coherent
light, when there is channel noise}). In order to check whether
the phase shifts induced by the phase shifters are in the chaotic
regime, we show in Fig.~\ref{Schematic of diagram of $M_1$, $M_2$
and $M_3$}(b) the Lyapunov exponents of Colpitts circuits with
different bandwidths. When the bandwidths of the Colpitts circuits
are smaller than $100$ MHz, the Lyapunov exponents of the Colpitts
circuits are equal $0$, which means that these circuits work in
the periodic regime. If we increase the bandwidths of the
circuits, the Colpitts circuits will then enter the chaotic regime
if the bandwidths are larger than $100$ MHz which corresponds to
positive Lyapunov exponents (see Fig.~\ref{Schematic of diagram of
$M_1$, $M_2$ and $M_3$}(b)). This is also confirmed by the phase
diagrams and the power spectra of the circuits with bandwidth of
$100$ MHz shown in Figs.~\ref{Schematic of diagram of $M_1$, $M_2$
and $M_3$}(c) and (d), and those of the circuits with bandwidth of
$500$ MHz in Figs.~\ref{Schematic of diagram of $M_1$, $M_2$ and
$M_3$}(e) and (f).

\begin{figure}[h]
\centerline{\includegraphics[width=8.6
cm,clip]{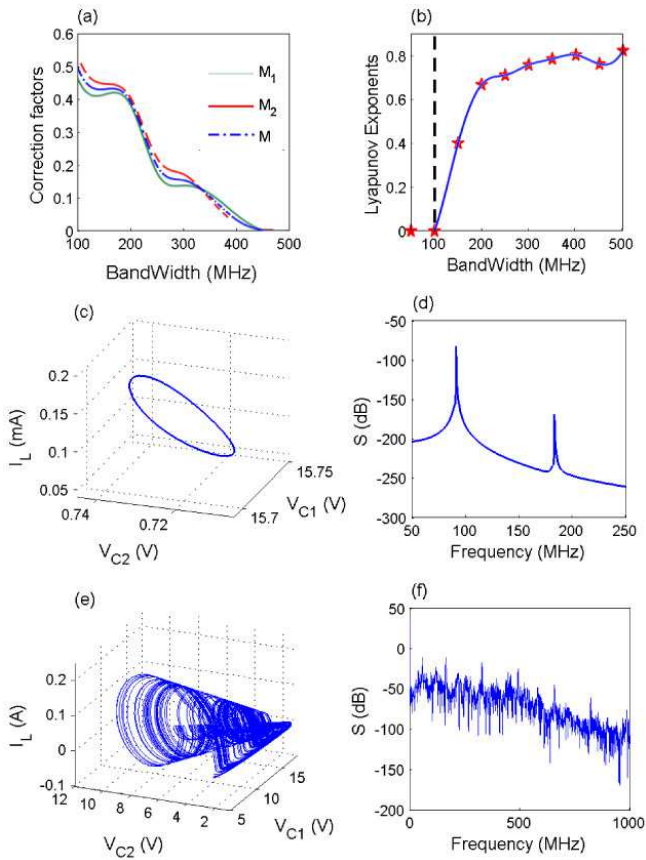}}\caption{(color online) (a) The factors $M_1$,
$M_2$, and $M$ versus the bandwidths of the Colitts circuits
without channel noise. The green solid curve represents the factor
$M_1$. The red dashed line denotes the curve for the factor $M_2$.
The blue dash-dot curve shows the factor $M=\sqrt{M_1M_2}$. (b)
The Lyapunov exponents of the Colpitts circuits versus different
bandwidths of the circuits. (c), (d) are the phase diagram and the
power spectrum of the Colpitts circuit with bandwidth $100$ MHz,
and (e), (f) correspond to the phase diagram and the power
spectrum of the Colpitts circuit with bandwidth of $500$
MHz.}\label{Schematic of diagram of $M_1$, $M_2$ and $M_3$}
\end{figure}

In order to show the efficiency of entanglement distribution by
the quantum CDMA network, we show in Fig.~\ref{Schematic of
diagram of fidelity $F_1$, $F_2$.} the fidelities $F_1=\langle
\Psi^+_{13}|\rho_{13}|\Psi^+_{13}\rangle$ and $F_2=\langle
\Psi^+_{24}|\rho_{24}|\Psi^+_{24}\rangle$ versus the bandwidths of
the chaotic signals and the average photon number of the probe
fields $\bar{n}$, where $|\Psi^+_{13}\rangle$
($|\Psi^+_{24}\rangle$) is the desired maximally-entangled state
between atoms $1$ and $3$ (atoms $2$ and $4$); and $\rho_{13}$
($\rho_{24}$) is the real state of the atoms $1$ and $3$ (atoms
$2$ and $4$). In our simulations, we take $\phi=\pi/3$, where
$\phi$ is the phase shift of the probe fields induced by the
qubit-cavity coupling. The trajectories of the fidelity $F_1$
versus the correction factor $M$ and the average photon number
$\bar{n}$ are given in Fig.~\ref{Schematic of diagram of fidelity
$F_1$, $F_2$.}(a). We can see clearly that the fidelity $F_1$ can
be very high if the factor $M$ is small enough and $\bar{n}$ is
not too small which corresponds to our previous analysis. If we
fix $\bar{n}=10$ and increase the bandwidth of the signals, we can
see the increase of the fidelities $F_1$ and $F_2$ as desired (see
Fig.~\ref{Schematic of diagram of fidelity $F_1$, $F_2$.}(b)).
Both $F_1$ and $F_2$ grow very quickly to approach $1$ with the
increase of the bandwidths of the signals to be larger than $400$
MHz, which corresponds to a perfect entanglement distribution.

\begin{figure}[h]
\centerline{\includegraphics[width=8
cm,clip]{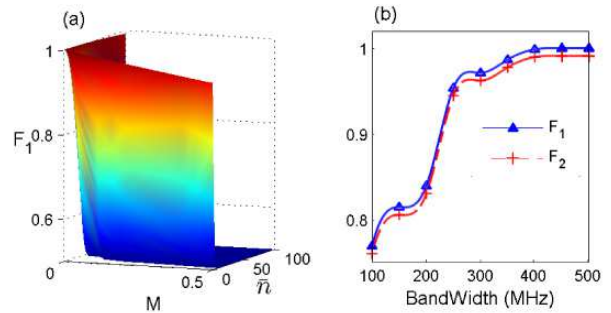}}\caption{(color online) (a) Fidelity $F_1$
versus different values of the correction factor $M$ and the
average photon number $\bar{n}$. (b) Trajectories of the
fidelities $F_1$ and $F_2$ versus different bandwidths of the
signals. Both $F_1$ and $F_2$ can be very close to the ideal case,
i.e., $F_1,\,F_2\approx1$ when the bandwidth of the signal is
larger than $400$ MHz, which means that we efficiently suppress
the interference effects of our quantum CDMA network induced by
the crosstalk between different data streams.}\label{Schematic of
diagram of fidelity $F_1$, $F_2$.}
\end{figure}

\section{NONIDEAL CASE: EFFECTS OF THE CHANNEL NOISE}\label{s4}

In the previous sections, we consider the ideal case in which the
channel noises are omitted. To show the efficiency of our method
in more practical case, we consider the effects of the channel
noises in this section~\cite{Jun1,Jun2}. The channel noises in
quantum communication may come from different sources, such as the
vibration of the optical fiber used for transmitting quantum
signals. Most of the channel noises, especially those induced by
the fibers, are low-frequency noises with several to several
hundreds of kHz, which is far smaller than the characteristic
frequency of the information-bearing fields, and also smaller than
the frequency band of the chaotic phase shifts introduced by the
chaotic circuits which is typically of several hundreds MHz. For
these reasons, we can omit the dynamical processes of the channel
noises and simply believe that they act as a beam splitter to
extract energy of the information-bearing field (see
Fig.~\ref{Schematic diagram of quantum CDMA network containing
channel noise}). As shown in Fig.~\ref{Schematic diagram of
quantum CDMA network containing channel noise}, the input-output
relationship of the beamsplitter ${\rm BS}_3$ used to represent
the effects of the channel noises can be written as
\begin{equation}\label{Input-output relationship of BS1 with channel noise}
a_7=\sqrt{1-\eta}a_5+\sqrt{\eta}a_{\rm NS},\quad
a_8=\sqrt{1-\eta}a_5-\sqrt{\eta}a_{\rm NS},
\end{equation}
where ${a_{\rm NS}}$ represents the noise mode and $\eta$ denotes
the decay rate induced by the noise. From Eqs.~(\ref{Input-output
relationship of the chaotic phase shifters}) to (\ref{Input-output
relationship between the senders and receiver: reduction}) and
Eq.~(\ref{Input-output relationship of BS1 with channel noise}),
we can obtain the input-output relationship of the noisy quantum
CDMA network as
\begin{eqnarray*}\label{Input-output relationship between the senders and receiver,when there is channel noise}
a_3&=&\frac{\sqrt{1-\eta}}{2}a_1+\frac{\sqrt{\left(1-\eta\right)M_1M_2}}{2}a_2+\sqrt{\frac{\eta M_1}{2}}a_{\rm NS}\\
&+&\sqrt{\frac{M_1}{2}}a_{\rm BS},\\
a_4&=&\frac{\sqrt{1-\eta}}{2}a_2+\frac{\sqrt{\left(1-\eta\right)M_1M_2}}{2}a_1+\sqrt{\frac{\eta M_2}{2}}a_{\rm NS}\\
&-&\sqrt{\frac{M_2}{2}}a_{\rm BS}.
\end{eqnarray*}
If we further assume that the field $a_{\rm NS}$ is in a vacuum
state, the output fields of the quantum CDMA network, i.e., $a_3$
and $a_4$, are in the following coherent states
\begin{eqnarray}\label{Output state of the quantum CDMA network under coherent light, when there is channel noise}
|\alpha_3\rangle &=&
\left|\frac{\sqrt{1-\eta}}{2}\alpha_1+\frac{\sqrt{\left(1-\eta\right)M_1M_2}}{2}\alpha_2\right\rangle,\nonumber\\
|\alpha_4\rangle &=& \left|\frac{\sqrt{1-\eta}}{2}\alpha_2+\frac{\sqrt{\left(1-\eta\right)M_1M_2}}{2}\alpha_1\right\rangle.
\end{eqnarray}

\begin{figure}[h]
\centerline{\includegraphics[width=8.6 cm,
clip]{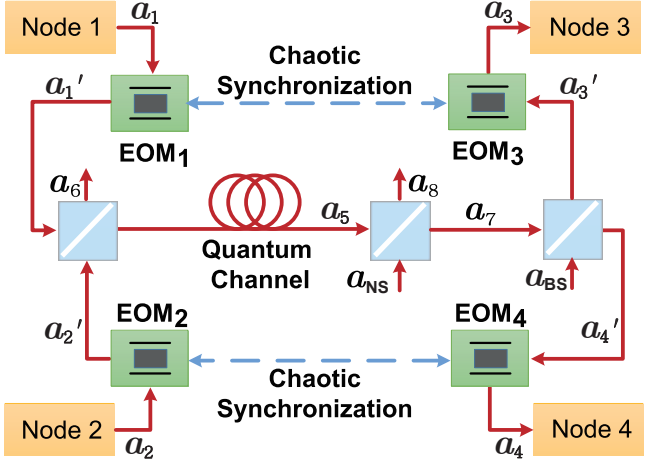}} \caption{(color online) Schematic diagram of
quantum CDMA network contains channel noise, where we use the beamsplitter ${\rm BS}_3$ to introduce channel noise.}\label{Schematic diagram of quantum CDMA network containing channel noise}
\end{figure}

\noindent Recall that the $i$-th atom is in the superposition
state
$|\psi_i\rangle=\left(|g_i\rangle+|e_i\rangle\right)/\sqrt{2}$,
and the probe field entering the cavity $1$ (cavity $2$) is a
bright coherent light $|\alpha\rangle$ with average photon number
$\bar{n}=|\alpha|^2\gg 1$. By omitting the terms related to
$\sqrt{M_1 M_2}$ in Eq.~(\ref{Output state of the quantum CDMA
network under coherent light, when there is channel noise}), we
can easily obtain the state of the total system composed of the
four atoms and output fields of the quantum network as
\begin{eqnarray*}\label{State of the total system when there is channel noise}
|\Psi\rangle&=&\left( \frac{1}{\sqrt{2}} | \Psi_{13}^{+}
\rangle \left| \frac{\sqrt{1-\eta}}{2} \alpha \right\rangle + \frac{1}{2} |
g_1 g_3 \rangle
\left| \frac{\sqrt{1-\eta}}{2} \alpha e^{-i \phi} \right\rangle \right. \nonumber \\
&& {\hspace{0cm}} \left. + \frac{1}{2} | e_1 e_3 \rangle \left|
\frac{\sqrt{1-\eta}}{2} \alpha e^{i \phi}
\right\rangle \right) \nonumber \\
&& {\hspace{0cm}} \left( \frac{1}{\sqrt{2}} \left| \Psi_{24}^{+}
\right\rangle \left| \frac{\sqrt{1-\eta}}{2} \alpha \right\rangle +\frac{1}{2}
| g_2 g_4 \rangle \left| \frac{\sqrt{1-\eta}}{2}\alpha e^{-i \phi}
\right\rangle
\right. \nonumber \\
&& {\hspace{0cm}} \left. +\frac{1}{2} | e_2 e_4 \rangle \left|
\frac{\sqrt{1-\eta}}{2} \alpha e^{i \phi} \right\rangle \right).
\end{eqnarray*}
If the decay rate $\eta$ induced by the channel noise is not quite
high and the probe fields are bright enough with average photon
number $\bar{n}=|\alpha|^2\gg1/\left(1-\eta\right)$, we have
\begin{eqnarray*}
&& \left|\left\langle {\frac{\sqrt{1-\eta}}{2} \alpha} \right. \left\vert {\frac{\sqrt{1-\eta}}{2} \alpha
e^{-i \phi}} \right\rangle\right|^2 =
e^{-(1-\eta)\bar{n}\sin^2\left(\phi/2\right)}
\approx 0, \\
&& \left| \left\langle {\frac{\sqrt{1-\eta}}{2} \alpha} \right. \left\vert {\frac{\sqrt{1-\eta}}{2} \alpha
e^{i \phi}} \right\rangle \right|^2 = e^{-(1-\eta)\bar{n}
\sin^2\left(
\phi/2 \right)} \approx 0, \\
&& \left| \left\langle {\frac{\sqrt{1-\eta}}{2} \alpha e^{-i \phi}} \right. \left\vert
{\frac{\sqrt{1-\eta}}{2} \alpha e^{i \phi}} \right\rangle \right|^2 = e^{
-(1-\eta)\bar{n}\sin^2\phi} \approx 0,
\end{eqnarray*}
which means that the three coherent states
$|\sqrt{1-\eta}\alpha/2\rangle$, $|\sqrt{1-\eta}\alpha
e^{-i\phi}/2\rangle$, and $|\sqrt{1-\eta}\alpha
e^{i\phi}/2\rangle$ are pairwise orthogonal and thus completely
distinguishable. Thus we can impose homodyne detections on the
fields leaking out of the cavities $3$ and $4$. If the
corresponding measurement outputs for the two probe fields are
both $\sqrt{1-\eta}\alpha/2$, the state of atoms $1$ and $3$ will
collapse to the maximally-entangled states $| \Psi_{13}^{+}
\rangle$ and that of the atoms $2$ and $4$ will collapse to the
maximally-entangled states $| \Psi_{24}^{+} \rangle$. From the
above discussions, we can conclude that our method is still valid
if only the decay rate induced by the channel noise is not quite
high such that the decayed probe fields are still bright enough.

However, if the decay rate $\eta$ induced by the channel noises
are too high such that the average photon number
$\bar{n}=|\alpha|^2$ is comparable to $1/\left(1-\eta\right)$, our
entanglement distribution strategy will not be so perfect. In this
case, we need to analyze the influence of noise on the fidelities
$F_1$ and $F_2$. Without loss of generality, let us focus on the
fidelity $F_1=\langle\Psi^+_{13}|\rho_{13}|\Psi^+_{13}\rangle$
versus different decay rate $\eta$ and average photon number
$\bar{n}$. The discussion for the fidelity $F_2$ are quite similar
and thus is omitted. We still choose $\phi=\pi/3$ and assume that
the correction factors $M_1=M_2\approx 0$. With these system
parameters, we show in Fig.~\ref{Schematic of diagram of fidelity
F1 considering channel noise} how the decay rate $\eta$ affects
the entanglement distribution. As can be seen from
Fig.~\ref{Schematic of diagram of fidelity F1 considering channel
noise}(a), fidelity $F_1$ can be very high when $\eta$ is small
and the average photon number is not large. Figure~\ref{Schematic
of diagram of fidelity F1 considering channel noise}(b) shows the
curves of the fidelity $F_1$ versus $\eta$ for several different
cases. The red solid curve represent the ideal case, i.e.,
$F_1=1$, which means that atoms $1$ and $3$ are in the maximally-entangled
state. The black dashed curve with plus signs denotes
the trajectory of the fidelity $F_1$ with increasing $\eta$
ranging from $0$ to $1$. The blue asterisks dash-dot curve shows
the fidelity $F_1$ versus $\eta$ without the four EOMs in our
quantum CDMA network. By comparing these three curves, we can see
that the fidelity $F_1$ will be greatly decreased if we move away
the four EOMs in our quantum CDMA network. In the meanwhile, with
EOMs, we can obtain a very high fidelity when the decay rate
$\eta$ is not too high.
\begin{figure}[h]
\centerline{\includegraphics[width=8
cm,clip]{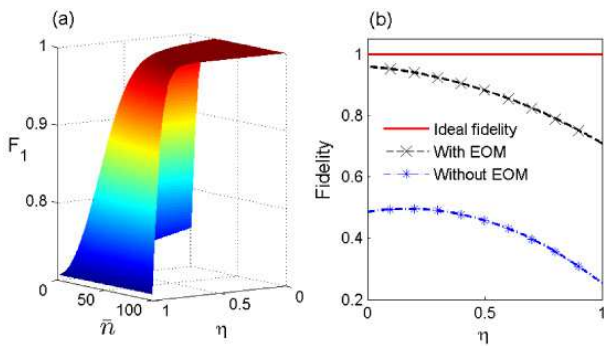}}\caption{(color online) (a) Fidelity $F_1$
versus different decay rates $\eta$ and the average photon number
$\bar{n}$. (b) Trajectories of $F_1$ versus $\eta$ with fixed
$\bar{n}=10$. The black dashed curve with plus signs shows the
curve of the fidelity $F_1$ realized by our quantum CDMA network.
The blue asterisks dash-dot curve represents the curve of $F_1$
when we the four EOMs are moved away. It is shown that our strategy can
still be valid when we consider the channel noise if only the
decay rate induced by the channel noises is not too
high.}\label{Schematic of diagram of fidelity F1 considering
channel noise}
\end{figure}

\section{CONCLUSIONS}\label{s5}
In summary, we present a strategy to distribute quantum
entanglement between two pairs of users via a single quantum
channel. The interference of the quantum signals from different
senders are greatly suppressed by introducing chaotic phase shifts
to broaden the quantum signals in the frequency domain. It is
shown that the two maximally-entangled states can be generated
between two pairs of nodes even when we consider the channel
noises. Our strategy is also hopeful to be applied to other
systems such as solid-state quantum circuits, and it also provides
new perspectives for the field of quantum network control.

\begin{center}
\textbf{ACKNOWLEDGMENTS}
\end{center}

CLZ would like to thank Dr. R.~B. Wu for helpful discussions.
JZ and YXL are supported by the National Basic Research
Program of China (973 Program) under Grant No. 2014CB921401, the
Tsinghua University Initiative Scientific Research Program, and
the Tsinghua National Laboratory for Information Science and
Technology (TNList) Cross-discipline Foundation. JZ is supported
by the NSFC under Grant Nos. 61174084, 61134008. YXL is supported
by the NSFC under Grant Nos. 10975080, 61025022, 91321208. FN is
partially supported by the RIKEN iTHES Project, MURI Center for
Dynamic Magneto-Optics, Grant-in-Aid for Scientific Research (A).
\\[0.2cm]

\section*{APPENDIX: CHAOTIC SYNCHRONIZATION OF COLPITTS OSCILLATOR CIRCUITS}\label{Chaotic synchronizationn of colpitts oscillator circuits}

In the system we consider, each pair of EOMs is driven by two
standard chaotic Colpitts oscillator circuits~\cite{Kennedy},
which are synchronized by the Pecora-Carroll synchronization
strategy~\cite{Shizhiguo,Parlitz}, as shown in Fig.~\ref{Schematic
diagram of the synchronized chaotic Colpitts circuit}. The
Colpitts chaotic synchronization circuit comprises of a
transmitter and a receiver. The transmitter is a standard Colpitts
oscillator circuit, which will enter the chaotic regime for
particular system parameters. The receiver is a part of the
standard Colpitts oscillator circuit. In our design, the system
parameters of the Colpitts oscillator circuits, such as the
resistance $R$, the inductance $L$, the capacitances $C_1$ and
$C_2$, and the voltage $V_{CC}$, are chosen as $R=27.99~\Omega$,
$L=17.5$~nH, $C_1=13.1$~pF, $C_2=12.7$~pF, and $V_{CC}=15$~V,
under which the synchronized voltages $V_{C2}$ and $\tilde{V}_{C2}$
are broadband chaotic signals with bandwidths of
$500$~MHz~\cite{Shizhiguo,Parlitz}.

\begin{figure}[h]
\centerline{\includegraphics[width=8.7 cm, clip]{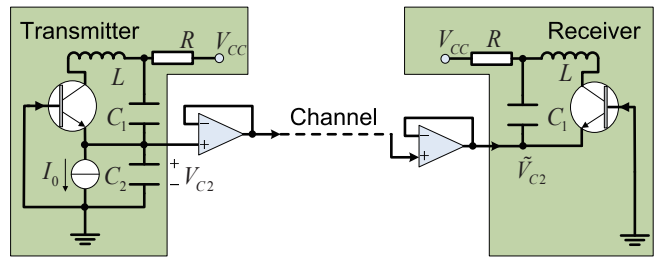}}
\caption{(color online) Schematic diagram of the synchronized
chaotic Colpitts circuits which is composed of a transmitter and a
receiver. Here we adopt the Pecora-Carroll synchronization
strategy to synchronize the two chaotic Colpitts
circuits~\cite{Pecora}.}\label{Schematic diagram of the
synchronized chaotic Colpitts circuit}
\end{figure}


\begin{thebibliography}{99}

\bibitem{Lukin} M. D. Lukin, Trapping and Manipulating Photon States in Atomic Ensembles,  Rev. Mod. Phys. {\bf 75}, 457 (2003). 

\bibitem{LMDuan} L.-M. Duan and C. Monroe, Quantum Networks with Trapped Ions, Rev. Mod. Phys. {\bf 82}, 1209 (2010). 

\bibitem{Sangouard} N. Sangouard, C. Simon, H. de Riedmatten, and N. Gisin, Quantum Repeaters Based on Atomic Ensembles and Linear Optics, Rev. Mod. Phys. {\bf 83}, 33 (2011). 

\bibitem{You} J.~Q. You and F.~Nori, Superconducting Circuits and Quantum Information, it Physics Today {\bf 58}, 42 (2005).

\bibitem{YMakhlin} Y. Makhlin, G. Sch\"{o}n, and A. Shnirman, Quantum-state Engineering with Josephson-junction Devices, Rev. Mod. Phys. {\bf 73}, 357 (2001). 

\bibitem{Xiang} Z.~L. Xiang, S. Ashhab, J.~Q. You, and F. Nori, Hybrid Quantum Circuits: Superconducting Circuits Interacting with Other Quantum Systems, Rev. Mod. Phys. {\bf 85}, 623 (2013). 

\bibitem{Kimble} H.~J. Kimble, The Quantum Internet, Nature {\bf 453}, 1023 (2008). 

\bibitem{Sillanpaa} M. A. Sillanp\"{a}\"{a}, J. I. Park, and R.~W. Simmonds, Coherent Quantum State Storage and Transfer Between Two Phase Qubits via a Resonant Cavity, Nature
    {\bf 449}, 438 (2007). 

\bibitem{Wang} X.-B. Wang, T. Hiroshima, A. Tomita, and M. Hayashi, Phys. Rep.
    {\bf 448}, 1 (2007).  

\bibitem{Pan} J.-W. Pan, Z.-B. Chen, C.-Y. Lu, H. Weinfurter, A. Zeilinger, and M. Zukowski, Multi-photon Entanglement and Interferometry, Rev. Mod. Phys. {\bf 84} 777 (2012). 

\bibitem{Gisin} N. Gisin, G. Ribordy, W. Tittel, and H. Zbinden, Quantum cryptography,  Rev. Mod. Phys. {\bf 74}, 145 (2002).  

\bibitem{Razavi} M. Razavi, Multiple-Access Quantum Key Distribution Networks, IEEE Trans. Commun. {\bf 60}, 3071 (2012).  

\bibitem{Mabuchi1} J.~I. Cirac, P. Zoller, H.~J. Kimble, and H. Mabuchi, Quantum State Transfer, and Entanglement Distribution among Distant Nodes in a Quantum Network,  Phys. Rev. Lett. {\bf 78}, 3221 (1997).  

\bibitem{Maitre} X. Maitre, E. Hagley, G. Nogues, C. Wunderlich, P. Goy, M. Brune, J. M. Raimond, and S. Haroche, Quantum Memory with a Single Photon in a Cavity, Phys. Rev. Lett. {\bf 79}, 769 (1997).  

\bibitem{Phillips} D.~F. Phillips, A. Fleischhauer, A. Mair, R. L. Walsworth, and M.~D. Lukin, Storage of Light in Atomic Vapor, Phys. Rev. Lett. {\bf 86}, 783 (2001).   

\bibitem{LMDuan2} L.~M. Duan, M.~D. Lukin, J.~I. Cirac, and P. Zoller, Long-distance Quantum Communication with Atomic Ensembles and Linear Optics, Nature {\bf 414}, 41 (2001).  

\bibitem{Matsukevich} D.~N. Matsukevich and A. Kuzmich, Quantum State Transfer Between Matter and Light, Science {\bf 306}, 663 (2004). 

\bibitem{Acin} A. Ac\'{i}n, J.~I. Cirac, and M. Lewenstei, Entanglement percolation
    in quantum networks, Nature Phys. {\bf 3}, 256 (2007).   

\bibitem{XinYou2} X.-Y. L\"{u}, J.-B. Liu, C.-L. Ding, and J.-H. Li, Dispersive Atom-field Interaction Scheme for Three-dimentional Entanglement Between Two Spatially Separated Atoms, Phys. Rev. A {\bf 78}, 032305 (2008).  

\bibitem{Felinto} D. Felinto, C. W. Chou, J. Laurat, E.~W. Schomburg, H. de Riedmatten, and H.~J. Kimble, Conditional Control of the Quantum States of Remote Atomic Memories for Quantum Networking, Nature Phys. {\bf 2}, 844 (2006). 

\bibitem{vanderWal} C. H. van der Wal, M.~D. Eisaman, A. Andre, R.~L. Walsworth, D.~F. Phillips, A.~S. Zibrov, M.~D. Lukin, Atomic Memory for Correlated Photon States,
    Science {\bf 301}, 196 (2003). 

\bibitem{Chou} C.-W. Chou, J. Laurat, H. Deng, K.~S. Choi, H. de Riedmatten, D. Felinto, and H.~J. Kimble, Functional Quantum Nodes for Entanglement Distribution over Scalable Quantum Networks, Science {\bf 316}, 316 (2007).  

\bibitem{Wilk} T. Wilk, S.~C. Webster, A. Kuhn, and G. Rempe, Single-Atom Single-Photon
    Quantum Interface, Science {\bf 317}, 488 (2007).  

\bibitem{Yard} G. Smith and J. Yard, Quantum Communication with Zero-Capacity Channels,
    Science {\bf 321}, 1812 (2008).  

\bibitem{Horodecki1} L. Czekaj and P. Horodecki, Purely Quantum Superadditivity of Classical Capacities of Quantum Multiple Access Channels, Phys. Rev. Lett. {\bf 102}, 110505 (2009). 

\bibitem{Horodecki2} M. Demianowicz and P. Horodecki, Quantum Channel Capacities: Multiparty
    Communication, Phys. Rev. A {\bf 74}, 042336 (2006).  

\bibitem{Rom} R. Rom and M. Sidi, Multiple Access Protocols: Performance and Analysis (Springer-Verlag, New York, 1990).  

\bibitem{Ipatov} V.~P. Ipatov, Spread Spectrum and CDMA: Principles and Applications (John Wiley \& Sons, Ltd, England, 2005).    

\bibitem{Frenzel} L.~E. Frenzel, Principles of Electronics Communications Systems (McGraw-Hill, New York, 2008).  

\bibitem{TMCover} T.~M. Cover and J.~A. Thomas,  Elements of Information Theory (Wiley,  New York, 1991), page 407. 

\bibitem{Godbout} G. Brassard, F. Bussieres, N. Godbout, and S. Lacroix, Multiuser quantum key distribution using wavelength division multiplexing, Proc. SPIE {\bf 5260},149 (2003). 

\bibitem{Townsend} P. Townsend, Simultaneous quantum cryptographic key distribution and conventional data transmission over installed fibre using wavelength-division multiplexing. Electron. Lett. {\bf 33}, 188 (1997).  

\bibitem{Yoshino} K. Yoshino, M. Fujiwara, A. Tanaka, S. Takahashi, Y. Nambu, A. Tomita, S. Miki, T. Yamashita, Z. Wang, M. Sasaki et al., High-speed Wavelength-division Multiplexing Quantum key Distribution System, Opt. Lett. {\bf 37}, 223 (2012).  

\bibitem{Brassard1} G. Brassard, F. Bussieres, N. Godbout, and S. Lacroix, Multiuser Quantum Key Distribution Using Wavelength Division Multiplexing, Proc. SPIE {\bf 5260}, 149 (2003).  

\bibitem{Brassard2} G. Brassard, F. Bussieres, N. Godbout, and S. Lacroix, Entanglement and Wavelength Division Multiplexing for Quantum Cryptography Networks, AIP Conf. Proc. {\bf 734}, 323 (2004).    

\bibitem{IChoiOE:2010} I. Choi, R.~J. Young, and P.~D. Townsend, Quantum key distribution on a 10Gb/s WDM-PON, Opt. Express {\bf 18}, 9600 (2010).   

\bibitem{JRoslundNatPhonon:2014} J. Roslund, R. Medeiros de Araujo, S. Jiang, C. Fabre,
    and N. Treps, Wavelength-multiplexed quantum networks with ultrafast frequency combs,  Nat. Photonics {\bf 8}, 109 (2014).    

\bibitem{AFurusawaNatPhotonics:2013} S. Yokoyama, R. Ukai, S.~C. Armstrong, C. Sornphiphatphong, T. Kaji, S. Suzuki, J. Yoshikawa, H. Yonezawa, N.~C. Menicucci, and A.
    Furusawa, Ultra-large-scale continuous-variable cluster states multiplexed in the time domain, Nat. Photonics {\bf 7}, 982 (2013).   

\bibitem{Fei} F. Li, L.~Z. Zhou, L. Liu, and H.~B. Li, A Quantum Search Based Signal Detection for MIMO-OFDM Systems, 18th Int. Conf. Telecommun., 276 (2011).  

\bibitem{Anandan} M. Anandan, S. Choudhary, and K.~P. Kumar, OFDM for Frequency Coded Quantum Key Distribution, Int. Conf. Fib. Optics. Photonics., 3 (2012).    

\bibitem{JZhang3} J. Zhang, Y.-X. Liu, S. K. \"{O}zdemir, R.-B. Wu, X.-B. Wang, F.~F.~Gao, X.~B. Wang, L. Yan, and F. Nori, Quantum Internet Using Code Division Multiple Access, Sci. Rep. {\bf 3}, 2211 (2013).  

\bibitem{Carcia} J.~C. Garcia-Escartin and P. Chamorro-Posada, Quantum Spread Spectrum Multiple Access, IEEE J. Sel. Top. Quantum. Electron. {\bf 21}, 6400107 (2015).   

\bibitem{Humble} T.~S. Humble, Quantum Spread Spectrum Communication, Quantum Information and Computation IX {\bf 8057}, 80570 (2011). 

\bibitem{Carcia2} J.~C. Garcia-Escartin and P. Chamorro-Posada, Quantum Multiplexing with Optical Coherent States, Quantum Inf. Comput. {\bf 9}, 573 (2009).  

\bibitem{Kennedy} M.~P. Kennedy, Chaos in The Colpitts Oscillator, IEEE Transactions on Circuits and Systems {\bf 41}, 771 (1994).   

\bibitem{Shizhiguo} Z.~G. Shi and L.~X. Ran, Microwave Chaotic Colpitts Oscillator: Design, Implementation and applications, J. of Electromagn. Waves and Appl.  {\bf 20}, 1335 (2006). 

\bibitem{Parlitz} U. Parlitz, L.~O. Chua, Lj. Kocarev, K.~S. Halle, and A. Shang, Transmission of Digital siganls by Chaotic Synchronization, Int. J. Bif. Chaos {\bf 2}, 973 (1992).  

\bibitem{Pecora} L.~M. Pecora and T.~L. Carroll, Synchronization in Chaotic Systems,  Phys. Rev. Lett. {\bf 64}, 821 (1990). 

\bibitem{Cuomo} K.~M. Cuomo and A.~V. Oppenheim, Circuit Implementation of Synchronized Chaos with Applications to Communications, Phys. Rev. Lett. {\bf 71}, 65 (1993).  

\bibitem{JZhang} J. Zhang, Y.-X. Liu, W.-M. Zhang, L.-A. Wu, R.-B. Wu, and T.-J. Tarn, Deterministic Chaos Can Act as a Decoherence Suppressor, Phys. Rev. B {\bf 84}, 214304 (2011).

\bibitem{Zhou} L. Zhou, S. Yang, Y.~X. Liu, C.~P. Sun, and F. Nori, Quantum Zeno Switch for Single-photon Coherent Transport, Phys. Rev. A {\bf 80}, 062109 (2009).  

\bibitem{Ashhab} S. Ashhab, J.~R. Johansson, A.~M. Zagoskin, and F. Nori, Two-level Systems Driven by Large-amplitude Fields, Phys. Rev. A {\bf 75}, 063414 (2007).  

\bibitem{Bergli} J. Bergli, Y.~M. Galperin, and B.~L. Altshuler, Decoherence in Qubits due to Low-frequency Noise,  New J. Phys. {\bf 11}, 025002 (2009).   

\bibitem{PvanLoock} P. van Loock, T.~D. Ladd, K. Sanaka, F. Yamaguchi, K. Nemoto, W.~J. Munro, and Y. Yamamoto, Hybrid Quantum Repeater Using Bright Coherent Light, Phys. Rev. Lett. {\bf 96}, 240501 (2006).   

\bibitem{Ladd} T.~D. Ladd, P. van Loock, K. Nemoto, W. J. Munro, and Y. Yamamoto, Hybrid Quantum Repeater Based on Dispersive CQED Interactions Between Matter Qubits and Bright Coherent Light, New J. Phys. {\bf 8}, 184 (2006).   

\bibitem{Blais2} A. Blais, R. S. Huang, A. Wallraff, S. M. Girvin, and R. J. Schoelkopf, Cavity Quantum Electrodynamics for Superconducting Electrical Circuits: An Architecture For Quantum Computation, Phys. Rev. A {\bf 69}, 062320 (2004).   

\bibitem{Kolchin} P. Kolchin, C. Belthangady, S.~W. Du, G.~Y. Yin, and S.~E. Harris, Electro-Optic Modulation of Single Photons, Phys. Rev. Lett. {\bf 101}, 103601 (2008).  

\bibitem{Tsang} M. Tsang, Cavity Quantum Electro-Optics, Phys. Rev. A {\bf 81}, 063837 (2010). 

\bibitem{Jun1} J. Jing and T. Yu, Non-Markovian Relaxation of a Three-Level System: Quantum Trajectory Approach, Phys. Rev. Lett. {\bf 105}, 240403 (2010).

\bibitem{Jun2} J. Jing, L.~A. Wu, M. Byrd, J.~Q. You, T. Yu, and Z.~M. Wang, Noperturbative Leakage Elimination Operators and Control of a Three-Level System, Phys. Rev. Lett. {\bf 114}, 190502 (2015).


\end{thebibliography}
\end{document}